\documentclass[pra,twocolumn,showpacs]{revtex4}
\usepackage{amsmath}
\usepackage{graphicx}

\begin{document}

\title{Relation between phase space coverage and entanglement for spin-1/2 systems}
\author{Stefan Schenk}
\author{Gert-Ludwig Ingold}
\affiliation{Institut f{\"u}r Physik, Universit{\"a}t Augsburg,
Universit{\"a}tsstra{\ss}e 1, D-86135 Augsburg, Germany}
\begin{abstract}
For systems of two and three spins 1/2 it is known that the second moment of the
Husimi function can be related to entanglement properties of the corresponding
states. Here, we generalize this relation to an arbitrary number of spins in
a pure state. It is shown that the second moment of the Husimi function can
be expressed in terms of the lengths of the concurrence vectors for all possible
partitions of the $N$-spin system in two subsystems. This relation implies that
the phase space distribution of an entangled state is less localized than that
of a non-entangled state. As an example, the second moment of the Husimi function 
is analyzed for an Ising chain subject to a magnetic field perpendicular to the 
chain axis.
\end{abstract}
\date{\today}

\maketitle

\section{Introduction}
In the last few years there has been growing activity in the study of the 
behavior of spin chains viewed from a quantum information perspective. The 
relations between condensed matter physics and quantum information are twofold
in this case. On the one hand, spin chains can provide a tool for quantum
communication \cite{bose03,burga05}. On the other hand, the concept of entanglement
has been employed to study spin systems, in particular at a quantum phase
transition \cite{oster02,osbor02,vidal03,lator04}. Furthermore, the concept
of matrix product states has led to new insights into the density matrix 
renormalization group algorithm (DMRG) with which the ground state properties
of spin systems can be determined \cite{verst04,schol05}.

\begin{figure}
\includegraphics[width=\columnwidth]{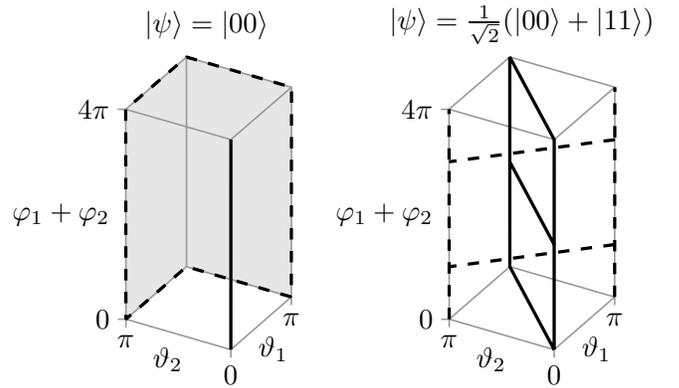}
\caption{\label{fig:twospin}Basic structure of the Husimi function of a 
factorizing state (left) and a Bell state (right). The full lines indicate the 
positions of the maxima of the Husimi function. Along the dashed lines and
on the gray planes the Husimi function vanishes. The entangled state has a more 
complex phase space structure than the non-entangled state. For details see 
Sec.~\ref{sec:phasespace}.}
\end{figure}

Recently, A.~Sugita \cite{sugit03} has pointed out for two spins 1/2 a 
relation between an entanglement measure, the so-called concurrence 
\cite{woott01}, and a property of the phase-space representation of the
state of the two spins. More specifically, this relation involves the second 
moment of the Husimi function, a positive definite phase space distribution. 
\cite{husim40,lee95} This quantity can be viewed as an inverse participation 
ratio in phase space and measures to which extent the phase space is covered 
by the Husimi function. It turns out that the more the state extends over
phase space, the more the state is entangled. In particular, the Husimi
function of a factorizable state minimally covers the phase space.
A first impression of this difference between the phase space representations
of entangled and non-entangled states can be obtained from Fig.~\ref{fig:twospin}.
In this figure, the basic structure of the Husimi function is visualized by
full lines where the maxima of the Husimi function are located. At dashed
lines and on gray planes the Husimi function vanishes. The comparison of a 
factorizing state on the left and a maximally entangled state on the right 
indicates that in the latter case the extension of the Husimi function is 
larger. The figure will be explained 
in more detail in Sec.~\ref{sec:phasespace} below where these first observations
will be made more precise.

For three spins an expression for the second moment of the Husimi function
was given by Sugita in terms of the concurrence between two of the three 
spins and the so-called 3-tangle \cite{coffm00}. As the Husimi function
can be defined for an arbitrary number of spins, it is 
interesting to determine whether and if yes how its second moment is 
related to entanglement measures for an arbitrary number of spins. 

Another motivation for this study arises from the fact that recently phase
space methods have been employed to analyze condensed matter systems. In 
particular the occurrence of a metal-insulator transition in disordered and
quasiperiodic systems has been analyzed by means of the inverse participation 
ratio in phase space \cite{wobst03,aulba04}. It appears natural to extend
these studies to interacting systems. Unfortunately, already the numerical
treatment of two particles moving in one dimension is quite demanding. Here,
spin systems like the one which we will discuss in Sec.~\ref{sec:ising} present 
significant advantages.  For the interpretation of the phase space properties,
it is again interesting to know the relation between the second moment of
the Husimi function and the entanglement of the state under consideration. 

The phase space approach has occasionally been used in the context of quantum
information in the past. The state and time evolution of a quantum computer
have been described by means of a discrete Wigner function \cite{mique02,paz05}.
More closely related to the present work is the investigation of the 
entanglement in bipartite systems on the basis of the Wehrl entropy where each 
subsystem is described in terms of a sufficiently large spin \cite{minte04a}. 
Furthermore, the relation between nonlinear bipartite systems and the
classical phase space dynamics has been considered \cite{hines05}.
In contrast, here we consider the phase space properties of an arbitrary number 
of distinguishable spins by means of a corresponding number of spin-1/2 
coherent states. A single number extracted from such a phase space 
representation can at best describe entanglement in a very global sense. 
Therefore, our intention is not to introduce a new quantity to 
measure entanglement but rather to provide a link between phase space 
properties and entanglement.

We start in Sec.~\ref{sec:phasespace} by reviewing the phase space representation for 
spin-1/2 systems and presenting Sugita's results in a form suitable for the ensuing 
discussion. In Sec.~\ref{sec:proj} the second moment of the Husimi function will
be expressed in terms of projectors acting on the Hilbert space and an auxiliary
copy. In this way, we make connection to the work by Mintert \textit{et al.} 
\cite{minte05} which allows us to obtain a relation to concurrences.
Finally, in Sec.~\ref{sec:ising} we discuss as an illustrative example the Ising 
spin chain in presence of a magnetic field perpendicular to the chain axis.

\section{Phase space for spins}
\label{sec:phasespace}
We will restrict our discussion to spin-1/2 systems where the spin-coherent
states, the analogue of the coherent states for the harmonic oscillator, are
given by \cite{radcl71,perel86}
\begin{equation}
\vert\vartheta,\varphi\rangle = 
\cos\left(\frac{\vartheta}{2}\right)\vert 0\rangle +
\sin\left(\frac{\vartheta}{2}\right)\mathrm{e}^{i\varphi}\vert 1\rangle
\label{eq:scs}
\end{equation}
so that each point on the Bloch sphere characterized by the two angles 
$\vartheta$ and $\varphi$ represents a spin-coherent state. We employ the 
notation common in quantum information, where $\vert 0\rangle$ and 
$\vert 1\rangle$ correspond to the eigenstates of the Pauli matrix $\sigma_z$
with eigenvalue $+1$ and $-1$, respectively. A coherent state for a system
consisting of $N$ distinguishable spins can be expressed as product of 
coherent states for each of the spins
\begin{equation}
\vert\mu\rangle = \prod_{i=1}^N\vert\vartheta_i,\varphi_i\rangle\,.
\label{eq:nscs}
\end{equation}

The spin-coherent states (\ref{eq:nscs}) enable us to define a positively 
definite phase space distribution
\begin{equation}
\mathcal{H}(\mu) = \langle\mu\vert\rho\vert\mu\rangle
\label{eq:h}
\end{equation}
which is the spin analogue of the Husimi- or Q-function familiar from the 
harmonic oscillator. \cite{husim40,lee95} In (\ref{eq:h}), $\rho$ denotes the 
density matrix of the state for which the phase space distribution is 
determined. 

In order to quantify the extension of a state in phase space, we introduce
the second moment of the Husimi function
\begin{equation}
\mathcal{P} = 3^N\int d\mu\, \mathcal{H}(\mu)^2
\label{eq:p}
\end{equation}
with the Haar measure $d\mu=\prod_{i=1}^N\sin(\vartheta_i)d\vartheta_i 
d\varphi_i/4\pi$. The prefactor is chosen such that a separable state leads
to $\mathcal{P}=1$. Up to a factor $(3/2)^N$, $\mathcal{P}$ is the inverse
participation ratio in phase space. Its inverse measures the extension of
the Husimi function in phase space. We note that the second moment of the
Husimi function corresponds to the first nontrivial term in the expansion 
of the Wehrl entropy $\int d\mu\mathcal{H}(\mu) \ln[\mathcal{H}(\mu)]$.
\cite{wehrl79}

To get a feeling for the physical content of the Husimi function
$\mathcal{H}$ and its second moment $\mathcal{P}$, we review results obtained 
for systems containing two or three distinguishable spins. For a pure two-spin 
state 
\begin{equation}
\vert\psi\rangle = a\vert 00\rangle + b\vert 01\rangle +c\vert 10\rangle +
d\vert 11\rangle
\label{eq:twospin}
\end{equation}
with $\vert a\vert^2+\vert b\vert^2+\vert c\vert^2+\vert d\vert^2=1$ the 
second moment of the Husimi function is obtained as
\begin{equation}
\mathcal{P} = 1 -\vert ad-bc\vert^2\,.
\label{eq:ptwo}
\end{equation}
For separable states, one has $ad=bc$ and therefore $\mathcal{P}=1$. On the 
other hand, the minimal value of the second moment of the Husimi function
for systems consisting of two spins is obtained for Bell states with 
$\mathcal{P}=3/4$. These results indicate the existence of a relation 
between this phase space quantity and the amount of entanglement. 

In order to illustrate the difference between the two cases, it is useful
to first consider the Husimi function which, for two spins 1/2, in general
is a function of the four angles $\vartheta_1, \vartheta_2, \varphi_1,$ and
$\varphi_2$. For the factorizing state $\vert\psi\rangle=\vert00\rangle$,
one obtains the Husimi function
\begin{equation}
\mathcal{H} = \frac{1}{4}\big(1+\cos(\vartheta_1)\big)
                         \big(1+\cos(\vartheta_2)\big)
\end{equation}
while for the Bell state $\vert\psi\rangle=(\vert00\rangle+\vert11\rangle)/
\sqrt{2}$ one finds
\begin{equation}
\mathcal{H} = \frac{1}{4}\big(1+\cos(\vartheta_1)\cos(\vartheta_2)+
                \sin(\vartheta_1)\sin(\vartheta_2)\cos(\varphi_1+\varphi_2)\big)\,.
\end{equation}
The fact that these Husimi functions depend only on three independent variables
allows us to represent their structure in Fig.~\ref{fig:twospin}.
For the factorizing state $\vert00\rangle$, the maximum depicted by a full
line lies at $\vartheta_1=\vartheta_2=0$ as should be expected on
the basis of (\ref{eq:scs}). The gray areas delimited by the dashed lines
indicate the planes at $\vartheta_1=\pi$ and $\vartheta_2=\pi$ where the
Husimi function vanishes. Similarly, for the Bell state in
Fig.~\ref{fig:twospin}b, the maxima at $\vartheta_1=\vartheta_2=0$ and $\pi$ 
indicate the presence of the states $\vert00\rangle$ and $\vert11\rangle$. 
The relative phase between the two states is encoded in the position of the 
bridges along $\vartheta_1= \vartheta_2$. The dashed lines again indicate 
zeroes of the Husimi function. We remark that the independence of the Husimi 
function on $\varphi_1-\varphi_2$ is specific to superpositions of the 
states $\vert00\rangle$ and $\vert11\rangle$.

The second moment of the Husimi function (\ref{eq:ptwo}) for two spins is related to
the concurrence \cite{hill97}
\begin{equation}
\mathcal{C} = \vert\langle\psi\vert\sigma_y\otimes\sigma_y\vert\psi^*\rangle\vert
\end{equation}
where the star denotes the complex conjugate in the eigenbasis of $\sigma_z$.
An alternative expression, which will be useful in the following, can be 
obtained by introducing an auxiliary Hilbert space with a copy of the state 
$\vert\psi\rangle$.  We denote the state in both Hilbert spaces as a column 
vector $\genfrac{\vert}{\rangle}{0pt}{1}{\psi}{\psi}$. Then the concurrence
becomes
\begin{equation}
\label{eq:conc_two}
\mathcal{C} = \left\vert\left(
\left\langle\begin{matrix}0\\1\end{matrix}\right\vert-
\left\langle\begin{matrix}1\\0\end{matrix}\right\vert
\right)
\left(
\left\langle\begin{matrix}0\\1\end{matrix}\right\vert-
\left\langle\begin{matrix}1\\0\end{matrix}\right\vert
\right)
\left\vert\begin{matrix}\psi\\\psi\end{matrix}\right\rangle
\right\vert
\end{equation}
where the first bracket operates in the Hilbert space of the first spin while
the second bracket operates in the Hilbert space of the second spin. A more
general discussion of expressing the concurrence in terms of projectors can
be found in Ref.~\onlinecite{minte05}. Beyond these formal considerations, such 
an auxiliary Hilbert space has very recently been employed to directly measure
the concurrence of two photons. \cite{walbo06}

For the general two-spin state (\ref{eq:twospin}), the concurrence is given by
\begin{equation}
\mathcal{C} = 2\vert b^*c^*-a^*d^*\vert
\label{eq:ctwo}
\end{equation}
so that by comparison of (\ref{eq:ptwo}) and (\ref{eq:ctwo}) one immediately 
obtains the relation
\begin{equation}
\mathcal{P} = 1-\frac{\mathcal{C}^2}{4}\,.
\label{eq:rel_two}
\end{equation}

For three spins, the second moment of the Husimi function can be
expressed in terms of the concurrences between two of the three spins $A$, 
$B$, and $C$ and the 3-tangle $\tau_{ABC}$ \cite{coffm00} as
\begin{equation}
\mathcal{P}=1 -\frac{1}{4}(\mathcal{C}_{AB}^2+\mathcal{C}_{AC}^2+
\mathcal{C}_{BC}^2)-\frac{3}{8}\tau_{ABC}\,.
\label{eq:rel_three}
\end{equation}
For the generalization to an arbitrary number of spins it is more suggestive 
to express this result in terms of the concurrences between one spin and the 
two others as
\begin{equation}
  \mathcal{P}=1 -\frac{1}{8}(\mathcal{C}_{A(BC)}^2+\mathcal{C}_{B(AC)}^2+
                             \mathcal{C}_{C(AB)}^2)\,.
\label{eq:rel_threea}
\end{equation}

We remark here that the cases of two and three spins are the simplest in
the sense that each partition into two subsystems will yield a single
concurrence. This is in general no longer true for more than three
spins, the case which we are going to address now. 

\section{Second moment of the Husimi function as a projection}
\label{sec:proj}

For the following considerations it is convenient to express the second moment
of the Husimi function (\ref{eq:p}) in terms of projectors onto symmetric and 
antisymmetric subspaces. We start by considering a system consisting of a 
single spin 1/2. The key idea is to express the square in the integrand of
(\ref{eq:p}) in terms of a tensor product of the spin Hilbert space and
an auxiliary copy of this Hilbert space. The similarity with the auxiliary
Hilbert space introduced in (\ref{eq:conc_two}) already hints at the possibility 
of a general relation between the second moment $\mathcal{P}$ and the 
concurrence $\mathcal{C}$. 

In order to distinguish between
tensor products of different spins which we note horizontally, the tensor
product between one spin Hilbert space and its auxiliary copy will be
denoted vertically. The density matrix $\varrho=\rho\otimes\rho$
refers to the tensor product of the density matrices in these two spaces.
The second moment of the Husimi function can then be written as
\begin{equation}
\mathcal{P} = \mathrm{Tr}\left(3\int d\mu\varrho\left\vert
\begin{matrix}\mu\\\mu\end{matrix}\right\rangle\left\langle\begin{matrix}
  \mu\\\mu\end{matrix}\right\vert\right)\,.
\end{equation}
By construction, the states $\genfrac{\vert}{\rangle}{0pt}{1}{\mu}{\mu}$
do not contain contributions antisymmetric under exchange of the two Hilbert
spaces.

Expressing the projectors in terms of the coherent states (\ref{eq:scs}), one 
can carry out the integrals over the angles $\vartheta$ and $\varphi$. It turns 
out that the second moment of the Husimi function can be expressed as 
\begin{equation}
\mathcal{P} = \mathrm{Tr}\left(\varrho P_s\right) = \langle P_s\rangle\,.
\label{eq:pproj}
\end{equation}
Here, $P_s=P_s^{(1)}+P_s^{(2)}+P_s^{(3)}$ 
with
\begin{gather}
P_s^{(1)}=\left\vert\begin{matrix}0\\0\end{matrix}\right\rangle
\left\langle\begin{matrix}0\\0\end{matrix}\right\vert,\quad
P_s^{(2)}=\left\vert\begin{matrix}1\\1\end{matrix}\right\rangle
\left\langle\begin{matrix}1\\1\end{matrix}\right\vert\notag\\
P_s^{(3)}=\frac{1}{2}\left(\left\vert\begin{matrix}0\\1\end{matrix}
\right\rangle+\left\vert\begin{matrix}1\\0\end{matrix}\right\rangle\right)
\left(\left\langle\begin{matrix}0\\1\end{matrix}
\right\vert+\left\langle\begin{matrix}1\\0\end{matrix}\right\vert\right)
\end{gather}
projects onto the symmetric eigenstates of two qubits while 
\begin{equation}
P_a=\openone-P_s =
\frac{1}{2}\left(\left\vert\begin{matrix}0\\1\end{matrix}\right\rangle-
\left\vert\begin{matrix}1\\0\end{matrix}\right\rangle\right)
\left(\left\langle\begin{matrix}0\\1\end{matrix}\right\vert-
\left\langle\begin{matrix}1\\0\end{matrix}\right\vert\right)
\end{equation}
projects onto the antisymmetric eigenstate.
Expectation values like in (\ref{eq:pproj}) are always to be understood in the 
extended space containing the original Hilbert space as well as a copy.

It is straightforward to generalize this reasoning to more than one qubit 
because a coherent state according to (\ref{eq:nscs}) is defined as a 
product of coherent states for each qubit. For $N$ qubits, the expression 
(\ref{eq:pproj}) then becomes
\begin{equation}
\mathcal{P} = \langle P_s^{\otimes N}\rangle\,.
\label{eq:pps}
\end{equation}
Making use of the decomposition $\openone = (P_s+P_a)^{\otimes N}$ we can
write this expression in the more complicated but useful form
\begin{equation}
\begin{aligned}
\mathcal{P} &= \left\langle\openone-(P_s+P_a)^{\otimes N}+P_s^{\otimes N}
\right\rangle\\
&= 1-\left\langle\left\{P_s^{\otimes N-1}\otimes P_a\right\}
+\left\{P_s^{\otimes N-2}\otimes P_a^{\otimes 2}\right\}\right.\\
&\qquad\qquad\qquad\left.
+\dots+\left\{P_a^{\otimes N}\right\}\right\rangle\,.
\end{aligned}
\label{eq:psproj}
\end{equation}
The curly braces imply a sum over all different orderings of projection 
operators. Noting that
\begin{equation}
  P_s-P_a = 
  \left\vert\begin{matrix}0\\0\end{matrix}\right\rangle
  \left\langle\begin{matrix}0\\0\end{matrix}\right\vert +
  \left\vert\begin{matrix}1\\1\end{matrix}\right\rangle
  \left\langle\begin{matrix}1\\1\end{matrix}\right\vert +
  \left\vert\begin{matrix}0\\1\end{matrix}\right\rangle
  \left\langle\begin{matrix}1\\0\end{matrix}\right\vert +
  \left\vert\begin{matrix}1\\0\end{matrix}\right\rangle
  \left\langle\begin{matrix}0\\1\end{matrix}\right\vert
\end{equation}
one can demonstrate the relation
\begin{equation}
  \text{Tr}\left(\varrho(P_s-P_a)^{\otimes N}\right) = \text{Tr}(\rho^2)
\end{equation}
which allows us to rewrite (\ref{eq:psproj}) as
\begin{equation}
\label{eq:pevenproj}
\begin{aligned}
  \mathcal{P} &= \frac{1}{2}(1+\text{Tr}(\rho^2))\\&\quad-
\left\langle\left\{P_s^{\otimes N-2}\otimes P_a^{\otimes 2}\right\}
+\left\{P_s^{\otimes N-4}\otimes P_a^{\otimes 4}\right\}
+\dots\right\rangle\,.
\end{aligned}
\end{equation}

From (\ref{eq:pevenproj}) it follows that for mixtures all terms in 
(\ref{eq:psproj}) will contribute while for pure states the terms with an 
odd number of projectors $P_a$ onto antisymmetric states are irrelevant. All 
these contributions clearly vanish because a state 
$\genfrac{\vert}{\rangle}{0pt}{1}{\psi}{\psi}$ is symmetric under exchange 
of the Hilbert space and its auxiliary copy. 

For the further discussion, we will restrict ourselves to pure states where
the second moment of the Husimi function now reads
\begin{equation}
\mathcal{P} = 1-
\left\langle\left\{P_s^{\otimes N-2}\otimes P_a^{\otimes 2}\right\}
+\left\{P_s^{\otimes N-4}\otimes P_a^{\otimes 4}\right\}
+\dots\right\rangle\,.
\label{eq:ppure}
\end{equation}
In particular, for at most three qubits only one term in the expectation value
will contribute, yielding the simple relations (\ref{eq:rel_two}) and
(\ref{eq:rel_three}). 

The expression (\ref{eq:ppure}) depends on a linear combination of projectors
with equal weight. It represents a special case of a class of operators which
can be employed to define a concurrence. Following Ref.~\onlinecite{minte05}, 
we introduce the $N$-partite concurrence of a pure state $\vert\psi\rangle$ 
describing a system consisting of $N$ spins
\begin{equation}
\label{eq:cn}
c_N(\psi) = 2^{1-N/2}\sqrt{(2^N-2)\langle\psi\vert\psi\rangle^2-
   \sum_i\text{Tr}\rho_i^2}
\end{equation}
and find for the second moment of the Husimi function
\begin{equation}
  \label{eq:pcn}
\mathcal{P} = 1 - \frac{c_N(\psi)^2}{4}\,.
\end{equation}
In (\ref{eq:cn}), $\rho_i$ is the reduced density matrix of a subsystem and the
sum runs over the $2^N-2$ subsystems containing at most $N-1$ spins. 
Alternatively, one can make use of the relation \cite{woott01,yu06} 
\begin{equation}
  c_N^2 = 2^{2-N}\bar{\mathcal{C}}^2
\end{equation}
where
\begin{equation}
	\bar{\mathcal{C}} = \left(\sum_\mathrm{partitions}\sum_\alpha 
\mathcal{C}_\alpha^2\right)^{1/2}
\end{equation}
describes the total length of the concurrence vectors for all partitions of
the system into two subsystems. The index $\alpha$ denotes the components
of the concurrence vector for a given partition. We thus arrive at the relation 
between the second moment of the Husimi function and the total length of the 
concurrence vectors
\begin{equation}
\label{eq:pconclen}
\mathcal{P} = 1-\frac{1}{2^N}\bar{\mathcal{C}}^2\,.
\end{equation}
This relation generalizes the results (\ref{eq:rel_two}) and (\ref{eq:rel_three})
which are immediately recovered by noting that for two and three spins each
concurrence vector contains only one component and that there exist one and three
partitions, respectively.

From (\ref{eq:pconclen}) one can conclude, that entanglement leads to a decrease
of the second moment of the Husimi function and thus to a larger spread of the
phase space distribution. The relevant measure of the entanglement here
is the total length of the concurrence vectors.

It is instructive to determine the second moment $\mathcal{P}$ for two different
entangled states, the $N$-qubit GHZ and W states. For the $N$-qubit GHZ state
\begin{equation}
\vert\text{GHZ}_N\rangle = \frac{1}{\sqrt{2}}(\vert00\cdots0\rangle 
+\vert11\cdots1\rangle)\,.
\label{eq:ghzn}
\end{equation}
one finds
\begin{equation}
\label{eq:pghz}
\mathcal{P}(\vert\text{GHZ}_N\rangle) = \frac{1}{2}+\frac{1}{2^N}\,.
\end{equation}
This result makes sense even for only one or two qubits. In the first case, one
finds $\mathcal{P}=1$ indicating a ``separable'' state while in the second case
the second moment of the Husimi function of a Bell state is recovered. As the 
number of qubits increases, $\mathcal{P}$ decreases and approaches the 
value of $1/2$ in the limit $N\to\infty$.

For the $N$-qubit W state
\begin{equation}
\begin{aligned}
\vert\text{W}_N\rangle &= \frac{1}{\sqrt{N}}(\vert100\cdots0\rangle +
\vert010\cdots0\rangle+\cdots)\\
&= \frac{1}{\sqrt{N}}\big[\vert1\rangle\otimes\vert00\cdots0\rangle + 
\vert0\rangle\otimes\sqrt{N-1}\vert\text{W}_{N-1}\rangle\big]
\end{aligned}
\label{eq:wn}
\end{equation}
one can derive the recursion relation
\begin{equation}
\mathcal{P}(\vert\text{W}_N\rangle)=\frac{(N-1)^2}{N^2} 
\mathcal{P}(\vert\text{W}_{N-1})+\frac{1}{N}\,.
\end{equation}
With the initial condition $\mathcal{P}(\vert\text{W}_1\rangle)=1$ one arrives
at the solution
\begin{equation}
\label{eq:pw}
\mathcal{P}(\vert\text{W}_N\rangle)=\frac{1}{2}+ \frac{1}{2N}\,.
\end{equation}
As for the GHZ state, the case of two qubits reproduces the second moment
of the Husimi function of a Bell state. With an increasing number of qubits, 
$\mathcal{P}$ decreases and approaches $1/2$ in the limit $N\to\infty$. However,
for all $N>2$, the second moment of the Husimi function of a W state is larger 
than that of a GHZ state. This reflects the fact that the concurrence $c_N$ for
a GHZ state is larger than that of a W state. \cite{carva04} Although $\mathcal{P}$
does not allow to distinguish the different kinds of entanglement present in the two
classes of states, a difference is nevertheless visible in the phase space structure 
which is more extended for a GHZ state.

From the results (\ref{eq:pghz}) and (\ref{eq:pw}) one might infer that $\mathcal{P}$
possesses a lower bound of $1/2$. This is however not the case. According to 
(\ref{eq:pps}), the second moment of the Husimi function for a system consisting of 
subsystems not entangled among each other is given by the product of the respective
$\mathcal{P}$'s of the subsystems. For $N$ pairs of spins in a Bell state, one finds
$\mathcal{P}=(3/4)^N$ which clearly goes to zero for $N\to\infty$.

\section{Ising model in a magnetic field}
\label{sec:ising}

In the last few years the connection between entanglement and quantum phase
transitions has been studied extensively, for example see 
Refs.~\onlinecite{oster02, osbor02, vidal03, lator04, gunly01, fazio05}. 
In most cases the concurrence has been used as a measure for bipartite
entanglement. Taking a different perspective we concentrate on the phase
space properties of such a transition. In the previous section we have seen how 
the concurrence is related to the second moment of the Husimi function. It
is therefore interesting to study this phase space quantity for a model 
exhibiting a quantum phase transition.

As an example we consider a one-dimensional chain of spins 1/2 described by the 
Hamiltonian
\begin{equation}
  \label{eq:hising}
  H = -J\sum_i\left[\sigma_i^z\sigma_{i+1}^z 
      +g\left(\sigma_i^z\cos\Theta+\sigma_i^x\sin\Theta\right)\right]\,.
\end{equation}
The first term for $J>0$ gives rise to a ferromagnetic coupling between
neighboring spins while the second term arises due to a magnetic field which 
is oriented at an angle $\Theta$ with respect to the $z$-axis. The parameter 
$g$ describes the ratio between the magnetic field strength and the interaction
strength between two neighboring spins. For the case of a transverse
magnetic field, $\Theta=\pi/2$, this model undergoes a quantum phase 
transition at $g=1$. \cite{sachd99}

\begin{figure}
\includegraphics[width=\columnwidth]{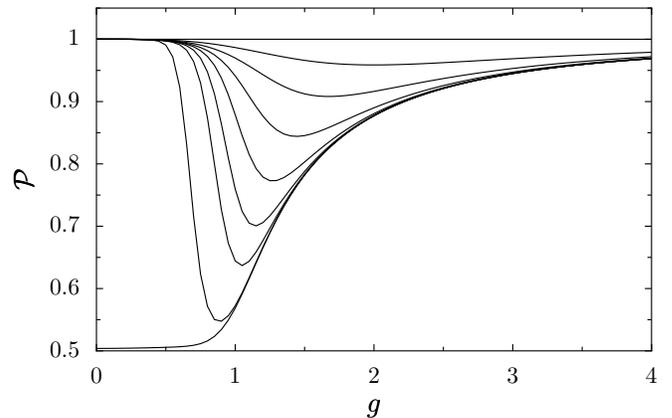}
\caption{\label{fig:Ising_8} The second moment $\mathcal{P}$ of the
Husimi function of a one-dimensional ferromagnetic Ising model with 8 spins 
and periodic boundary conditions
is shown as a function of the coupling constant $g$ for various angles 
between the $z$-axis and the magnetic field taking the values $\Theta/\pi= 
0, 0.42, 0.46, 0.48, 0.49, 0.495, 0.4975, 0.4995,$ and $0.5$ from the upper 
to the lower curve.}
\end{figure}

For $\Theta=0$, the ground state of (\ref{eq:hising}) will be given by a 
factorizing state either with all spins in state $\vert0\rangle$ or
$\vert1\rangle$ depending on the sign of $g$. Then, $\mathcal{P}=1$ 
independent of $g$. For angles $0<\Theta<\pi/2$ and $g\ll1$, the spins will
mostly be in the factorizing state $\vert0\dots0\rangle$. However, as $g$
increases, entanglement is build up and the extension of the state in phase
spaces increases. Correspondingly, $\mathcal{P}$ will decrease. On the
other hand, for $g\gg1$ the ferromagnetic coupling becomes irrelevant.
Then, all spins point in the direction of the magnetic field and $\mathcal{P}$ 
should reach an asymptotic value of 1. For a transverse magnetic field, 
$\Theta=\pi/2$, the ground state in the thermodynamic limit, $N\to\infty$, 
for $g<1$ will be a GHZ state. According to (\ref{eq:pghz}), we expect 
$\mathcal{P}=1/2$. On the other hand, for $g\gg1$, the asymptotic value
$\mathcal{P}=1$ should again be reached.

In Fig.~\ref{fig:Ising_8}, we present numerical results for a system of 8
spins and angles $\Theta$ varying from $0$ to $\pi/2$. From the numerically
obtained ground state of (\ref{eq:hising}), the second moment of the Husimi
function has been determined by evaluation of (\ref{eq:pps}). The lowest
curve represents the case of the transverse Ising model. For $g\ll1$, one
finds a value for $\mathcal{P}$ very close to $1/2$ as expected from
(\ref{eq:pghz}). For a finite number of spins, $\mathcal{P}$ displays a slight
increase even below $g=1$ where in the thermodynamic limit $\mathcal{P}$ is
expected to remain at a value of 1/2 before increasing for $g>1$ and reaching
$\mathcal{P}=1$ asymptotically. Even for angles $0<\Theta<\pi/2$, the curves 
in Fig.~\ref{fig:Ising_8} clearly show that entanglement is built up around 
$g=1$. The precursor of the quantum phase transition thus manifests itself also 
in the phase space properties for angles close to $\pi/2$.

We remark that while here we have focussed on the ground state properties,
the second moment of the Husimi function can also be determined at finite
temperatures. Extensions to the antiferromagnetic coupling and the 
two-dimensional Ising model are possible within the limits imposed by finite 
computer resources. \cite{schen05}

\section{Conclusions}
The relation between the second moment of the Husimi function and the
concurrence derived in \cite{sugit03} for two and three spins has been
generalized to an arbitrary number of spins. The extension of a state 
in phase space is thus related to its global entanglement properties.
Generally, entanglement will imply a delocalization of the Husimi function 
of a state. Furthermore, the result (\ref{eq:pcn}) provides a phase space
interpretation of the $N$-partite concurrence (\ref{eq:cn}).

The relation between phase space properties and entanglement has been
illustrated by calculating the second moment of the Husimi function for the 
one-dimensional Ising model with magnetic field. The precursor of a quantum 
phase transition is clearly seen in the presence of a transverse magnetic 
field even for a relatively small number of spins. For fields deviating from 
the transverse direction, the build-up of entanglement close to the critical 
value of the field can still be observed in the phase space properties.

\end{document}